\documentclass[11pt]{iopart}

\usepackage[british]{babel}
\usepackage{color}
\usepackage{mathrsfs}
\usepackage{setstack}
\usepackage{iopams}
\usepackage[T1]{fontenc}
\usepackage{graphicx}
\usepackage{color}

\newcommand{\xm}{x_{\mathrm{max}}}
\newcommand{\vm}{v_{\mathrm{max}}}
\newcommand{\mod}{\mathrm{mod}}

\newcommand{\rev}[1]{\textcolor{black}{#1}}

\begin{document}

\title{Conservative random walks in confining potentials}

\author{Bart{\l}omiej Dybiec$^\dagger$, Karol Capa{\l}a$^\dagger$, Aleksei V.
Chechkin$^{\ddagger,\sharp}$, and Ralf Metzler$^\sharp$}

\address{$\dagger$ Marian Smoluchowski Institute of Physics,
and Mark Kac Center for Complex Systems Research, Jagiellonian University, ul.
St. {\L}ojasiewicza 11, 30--348 Krak\'ow, Poland}
\address{$\ddagger$ Akhiezer Institute for Theoretical Physics NSC KIPT,
Kharkov 61108, Ukraine}
\address{$\sharp$ Institute for Physics \& Astronomy, University of Potsdam, 14476
Potsdam-Golm, Germany}
\ead{bartek@th.if.uj.edu.pl}

\begin{abstract}
L\'evy walks are continuous time random walks with spatio-temporal coupling of
jump lengths and waiting times, often used to model superdiffusive spreading
processes such as animals searching for food, tracer motion in weakly chaotic
systems, or even the dynamics in quantum systems such as cold atoms. In the
simplest version L\'evy walks move with a finite speed. Here, we present an
extension of the L\'evy walk scenario for the case when external force fields
influence the motion. The resulting motion is a combination of the response to
the deterministic force acting on the particle, changing its velocity according
to the principle of total energy conservation, and random velocity reversals
governed by the distribution of waiting times. For the fact that the motion stays
conservative, that is, on a constant energy surface, our scenario is
fundamentally different from thermal motion in the same external potentials.
In particular, we present results for the velocity and position distributions
for single well potentials of different steepness. The observed dynamics with
its continuous velocity changes enriches the theory of L\'evy walk processes
and will be of use in a variety of systems, for which the particles are
externally confined.
\end{abstract}

\section{Introduction}
\label{sec:intr}

Already in 1795 Dutch physician Jan Ingenhousz observed irregular motion of coal
dust particles on the surface of alcohol. Scottish botanist Robert Brown used
more systematic studies of jittery motion of various inanimate materials
following his observation of the same zigzagging of pollen granules extracted
from pollen grains, in 1828 \cite{brown1828}. More systematic studies of
diffusive particle motion are due to Gouy \cite{gouy1888note} and, in particular, Perrin,
whose seminal 1908 paper \cite{perrin1908agitation} prompted a whole series of even more
refined experiments. Two remarkable examples of the latter are the experiments
of Nordlund \cite{nordlund1914neue} and Kappler \cite{kappler1931versuche} using moving film plates
to produce long, individual time series of the motion. The theoretical analysis
of the Brownian motion of thermally activated particles appeared at the start
of the 20th century, mainly promoted by Smoluchowski \cite{smoluchowski1906}, Einstein \cite{einstein1905}, and Langevin \cite{langevin1908theorie}.

A standard way to define Brownian motion within the theory of stochastic processes
is in terms of the Wiener process, which can be represented as the Langevin equation
\cite{cox1965,risken1984}
\begin{equation}
\label{eq:wiener}
\frac{dx(t)}{dt}=\xi(t),
\end{equation}
where the Gaussian white noise $\xi(t)$  with zero mean $\langle\xi(t)\rangle=0$
and $\delta$-correlation $\langle\xi(t)\xi(t')\rangle=\delta(t-t')$ correspond to
the increments of the Wiener process. The latter are independent and identically
distributed according to the normal (Gaussian) distribution $x(t)-x(s)\sim\mathcal{
N}(0,t-s)$. Physically, this is the natural consequence of the assumption that
interactions of the test particle with its thermal environment occur on a faster
time scale. Moreover they are independent and bounded. Therefore, the collisions
with the bath particles can be approximated by Gaussian white noise. Note that
here and in the following we adopted a dimensionless notation, in which position,
time, and mass have unit dimension.

The Wiener process can be extended to $\alpha$-stable motions, for which the
increments remain independent but are distributed according to $\alpha$-stable
densities with heavy-tailed power-law asymptotics (L{\'e}vy flights)
\cite{samorodnitsky1994,janicki1994}. The Wiener process is included in the
$\alpha$-stable processes in the limiting case $\alpha=2$. $\alpha$-stable processes
can be extended to the case of external forces, resulting in the more general
form of the Langevin equation
\begin{equation}
\frac{dx(t)}{dt}=-V'(x)+\xi(t),
\label{eq:langevin}
\end{equation}
where $V(x)$ is the potential resulting in the deterministic force $F(x)=-V'(x)$.
Equation (\ref{eq:langevin}) provides a starting point for the analysis of a large
variety of noise induced phenomena \cite{horsthemke1984,shlesinger1995,fogedby1994b,
jespersen1999}. We
note that L{\'e}vy flights are often invoked as optimal search strategies, due to
their fractal sample paths combining longer excursions with local search events
\cite{viswanathan2011physics}. However in the presence of an external drift
the advantage of L{\'e}vy flights over normal Brownian search may become
significantly reduced or even turn into a disadvantage \cite{palyulin2014}.

The stochastic equation (\ref{eq:langevin}) displays some non-physical properties.
These are often negligible on the relevant space and time scales. However, from
a conceptual point of view random walks generated by the scheme (\ref{eq:langevin})
involve an infinite propagation speed. For Brownian motion this shortcoming was
particularly realised in the context of heat flow. There it was remedied by the
introduction of the Cattaneo or telegrapher's equation on the level of the
diffusion equation, leading to short-time ballistic motion with a finite horizon
of propagation \cite{cattaneo1948sulla,jou1996extended,dejagher1980hyperbolic}.
Even more so this problem arises for
$\alpha$-stable noises with $\alpha < 2$: due to the significant probability of
extremely long jumps the variance diverges \cite{montroll1984,metzler2000,me2000}.
A remedy for this infinite propagation speed can be introduced in terms of a
spatio-temporal coupling of jump lengths and associated waiting times. This concept
was introduced in the continuous time random walk scheme in the form of L{\'e}vy
walks by Shlesinger and coworkers in 1982 \cite{shlesinger1982}, see also
\cite{shlesinger1986,zaburdaev2015levy}.

In their simplest version L\'evy walks couple jump lengths and the
corresponding waiting time by a constant speed $v$. Random velocity
changes occur after independent and identically distributed waiting
times. These waiting times thus determine the flight time and thus
the travelled distance in between velocity changes. L\'evy walks were
successfully applied to model the dynamics of tracer particles in weakly
chaotic systems \cite{zumofen1993power,geisel1984anomalous,solomon1993} and
in the dynamics of cold atom systems \cite{kessler2012theoryfractional}. A
particular field of application of L\'evy walks are random search
processes, for instance, of animals searching for sparse food. Here
L\'evy walks combine the above-mentioned advantage of L{\'e}vy
flights with a physically meaningful, finite variance of the motion
\cite{shlesinger1986,lomholt2008levystrategies}. Remarkably, L\'evy
walk search statistics were unveiled in intracellular motion driven
by molecular motors \cite{song2018neuronal,chen2015memoryless}.
We also note that L\'evy walks exhibit ultraweak ergodicity breaking
such that time and ensemble averaged observables merely differ by a
constant, $\alpha$-dependent factor, and that they fulfil a linear
response relation, among other interesting physical properties
\cite{metzler2014anomalous,godec2013finite,godelc2013linear,
froemberg2013random,froemberg2015asymptotic}.

In the classical L\'evy walk setup \cite{zaburdaev2015levy} no forces are acting
on the moving particle. Therefore, its energy $\mathcal{E}$ is fixed to the kinetic
energy $v^2/2$, and is constant. At present, despite the wide use of the L{\'e}vy
walk model a comprehensive conceptual understanding of the response of L{\'e}vy
walks to an arbitrary external force field remains elusive. Here, we present a
possible extension of the L\'evy walk framework which takes into account such
force fields. As in the classical L\'evy walk model the system is considered to
be conservative in the sense that the total
energy is conserved, while the velocity is randomly reversed at time events defined
by the waiting time distribution. This conservative scheme immediately implies
that the L\'evy walk speed is no longer constant but constantly varies along
with the potential energy corresponding to a given position $x$. The resulting
stationary distribution will therefore differ from the Boltzmann distribution,
which emer\-ges in a thermal system and allows for infinitely far yet exponentially
suppressed excursions. The conservative model with well-defined maximum
excursions due to the constant energy requirement will thus be more appropriate
for cases when the particle is not allowed to cross a maximum distance while
experiencing a restoring force. An example for such a behaviour could be the
confinement of an animal searching for food while being limited to move only
within their homing range. When approaching the border of the homing range the
animal becomes more reluctant and thus slower. In such a scenario it may thus
make sense that the velocity decreases with the distance from the centre of the
system. More microscopically our conservative L{\'e}vy walk model may represent
a molecular motor in a biological cell \cite{song2018neuronal,chen2015memoryless}
attempting to pull an anchored cargo. Indeed, with growing resisting force molecular
motors achieve lower speeds and efficiencies \cite{julicher,goychuk}.

In the following section \ref{sec:model} we introduce our conservative L\'evy walk
model. Its properties are explored in detail in sections \ref{trajs} and
\ref{sec:results}. Finally we draw our conclusions in section \ref{sec:summary}.

\section{Conservative random walk model}
\label{sec:model}

\begin{figure}
\centering
\includegraphics[width=10cm]{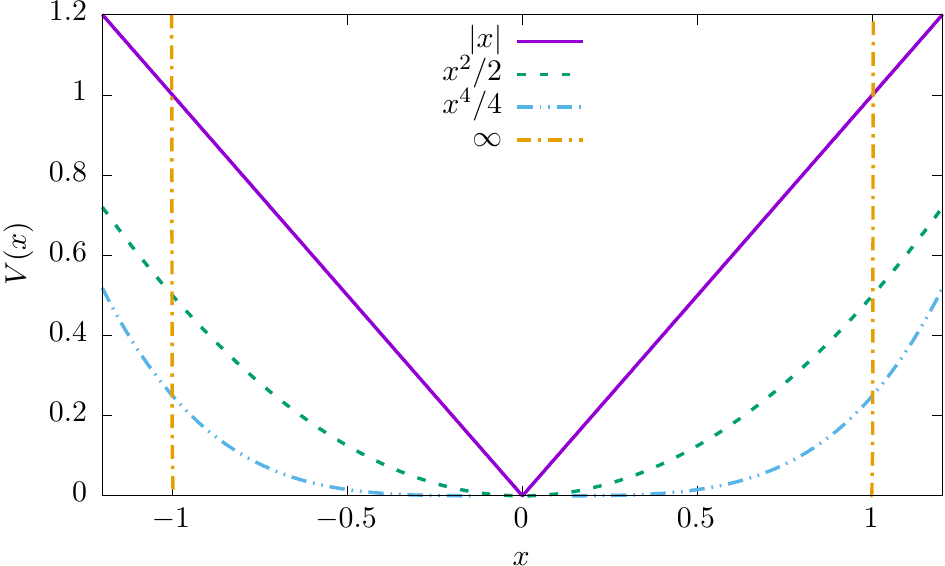}
\caption{Single well potential $V(x)=|x|^n/n$ for $n=1,2,4$ as well as for
$n=\infty$, which represents an infinite rectangular potential well.}
\label{fig:potential}
\end{figure}

\begin{figure}
\centering
\includegraphics[width=10cm]{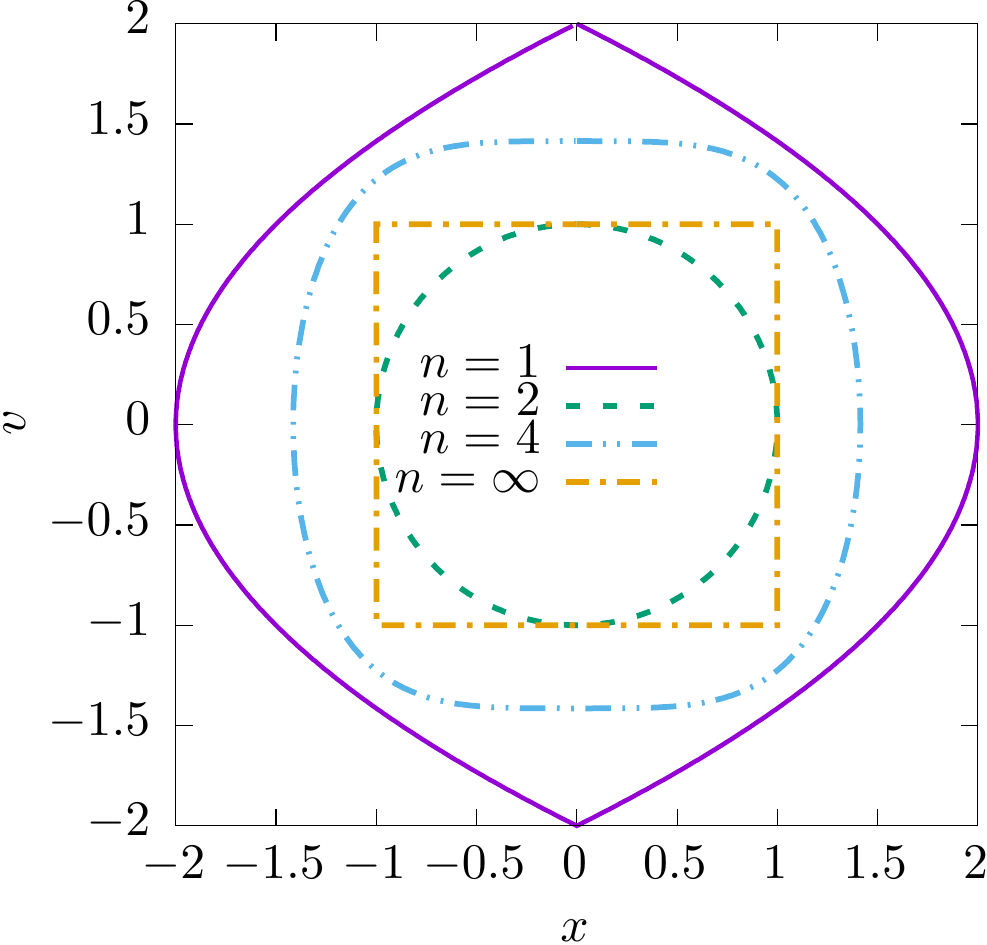}
\caption{Phase space portrait of constant energy curves for $n=1,2,4$,
and $\infty$. Note the soft velocity reversals at the points of maximal
distance from the origin, where $v(\xm)=0$.}
\label{fig:xv}
\end{figure}

We start with the motion of a test particle described by the classical Newton
equation
\begin{equation}
\frac{d^2x(t)}{dt^2}=-V'(x),
\label{eq:newton}
\end{equation}
in our dimensionless notation. For the external potential we choose the power-law
form
\begin{equation}
V(x)=\frac{|x|^n}{n},
\label{eq:potential}
\end{equation}
for which we consider the integer values $n=1,2,4$ and $n=\infty$, that is,
a softer-than-harmonic, harmonic, and stronger-than-harmonic potential as well
as a box with infinitely steep walls, see figure \ref{fig:potential}. The motion
encoded by equation (\ref{eq:newton}) for confining potentials of the form
(\ref{eq:potential}) is always periodic with the period
\begin{equation}
T=4\int_0^{\xm}\frac{dx}{\sqrt{2[\mathcal{E}-V(x)]}}.
\label{eq:period}
\end{equation}
Here $\xm$ is the reversal point satisfying $V(\xm)=\mathcal{E}$, and thus the
period of the motion depends on the system energy $\mathcal{E}$, except for the
harmonic case $n=2$, see reference \cite{landau1988theoretical}. The total energy
$\mathcal{E}$ is determined by the initial condition $x(0)$ and $\dot{x}(0)$, or,
more precisely $\mathcal{E}=\frac{1}{2}\dot{x}^2(0)+\frac{1}{n}|x(0)|^n$. In phase
space, a
particle described by relation (\ref{eq:newton}) moves (clockwise) along the
closed orbit determined by the system energy $\mathcal{E}$
\begin{equation}
\mathcal{E}=\frac{1}{2}v^2+V(x)=\mathrm{const},
\label{eq:conservation}
\end{equation}
see figure \ref{fig:xv}. Naturally velocity reversals occur at the points of
maximal distance from the origin $\xm$. These reversals are \emph{soft\/} in the sense
that they occur at $v(\xm)=0$.

In addition to the naturally occurring velocity reversals in the orbits of
figure \ref{fig:xv} we now introduce additional, random velocity reversals.
In this scheme a particle moves according to equation (\ref{eq:newton}) for
a random time $\tau$ ($\tau\geqslant0$) which is distributed according to the
given density $\psi(\tau)$ of waiting times. With the probability $p=1$, the
velocity is reversed, that is, $v(\tau)\to-v(\tau)$. These reversals are
\emph{hard\/}
in the sense that they typically occur at points where $v(x)\neq0$. Here, we assume that
$\tau=|\zeta|$, where $\zeta$ follows a symmetric $\alpha$-stable density
with characteristic function \cite{samorodnitsky1994}
\begin{equation}
\phi(k)=\langle\exp(ik\tau)\rangle=\exp\left(-\kappa^\alpha|k|^\alpha\right),
\label{eq:charact}
\end{equation}
where $\alpha$ ($0<\alpha\leqslant2$) is the stability index and $\kappa$
($\kappa>0$) is the scale parameter.\footnote{Note that we could have chosen
a completely asymmetric, one-sided L{\'e}vy stable distribution instead.
However, with the symmetric choice we can also consider $\alpha$ values
larger than unity \cite{samorodnitsky1994,hughes}.} For $\alpha<2$, $\psi(\tau)$
asymptotically behaves like the power-law
\begin{equation}
\psi(\tau)\simeq\frac{1}{\tau^{\alpha+1}}.
\label{eq:tauasymptotics}
\end{equation}
Consequently, for $\alpha<1$, the average between reversal time diverges.
Additionally, we will consider the exponential waiting time distribution
(Brownian creepers \cite{campos2015optimal})
\begin{equation}
\psi(\tau)=\lambda\exp(-\lambda\tau)
\label{eq:expo}
\end{equation}
with width $\lambda$. In this latter case moments of any order are finite.

\begin{figure}
\centering
\includegraphics[width=10cm]{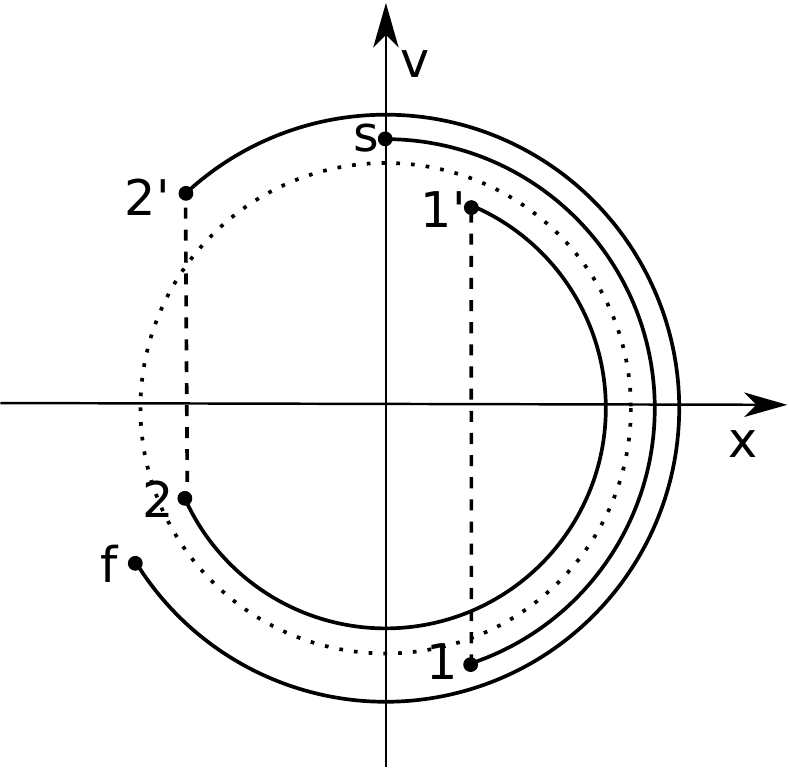}
\caption{Schematic of the particle motion in phase space from starting point
(s) to final point (f). Solid line shows a possible trajectory while dashed
parts correspond to velocity reversals ($1\to 1'$) and ($2\to 2'$). The dotted
circle represents $\mathcal{E}=\mathrm{const}$ orbit on which a particle stays
all the time, the different radii are only drawn for clarity.}
\label{fig:phasespace}
\end{figure}

\begin{figure*}
\centering
\includegraphics[width=12.8cm]{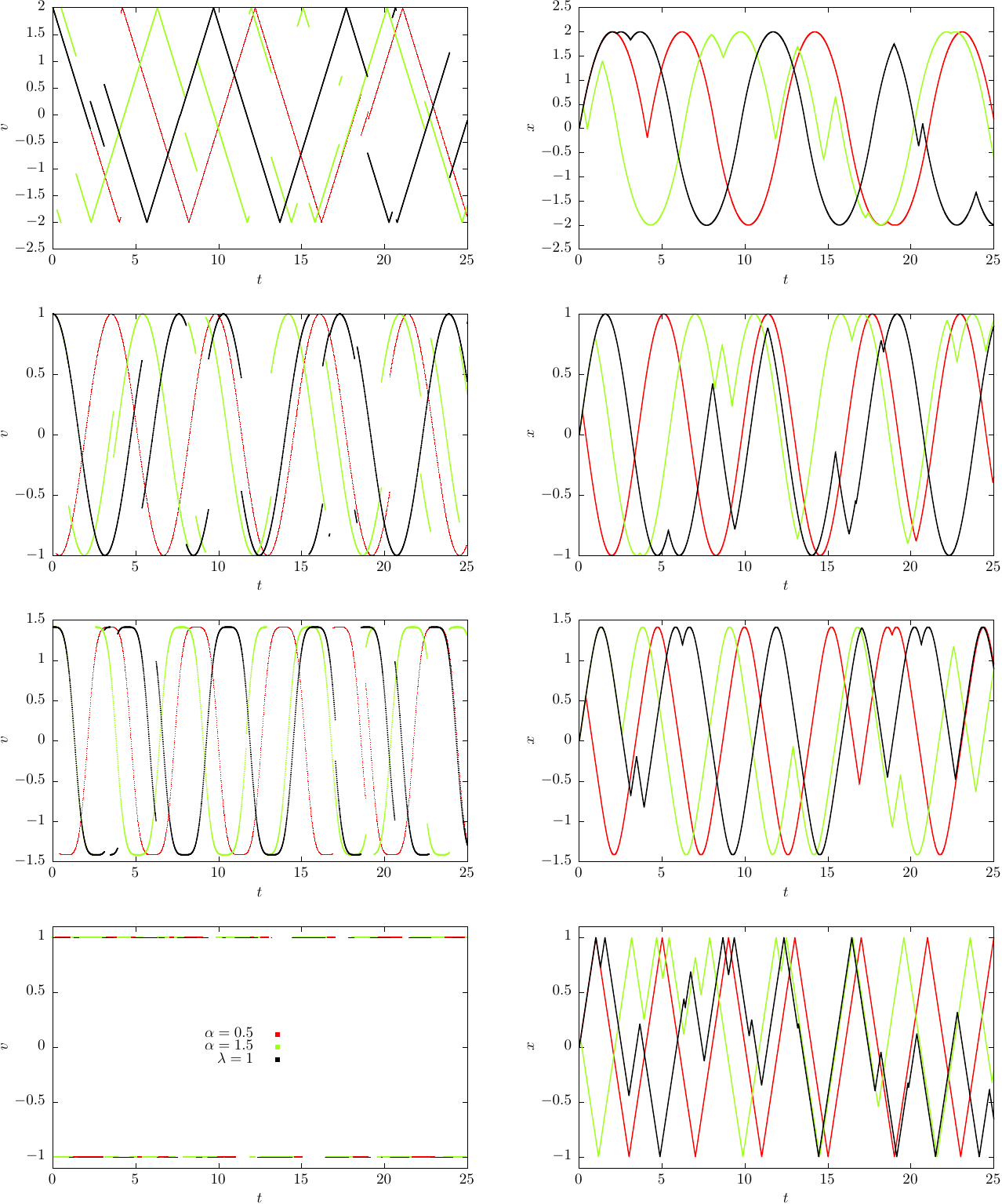}
\caption{Sample trajectories for $n=1$, $2$, $4$, and $\infty$ (top to bottom).
The different colours correspond to $\alpha=0.5$ (red), $\alpha=1.5$ (green),
and the exponential case with the scale factor $\lambda=1$ (black), see bottom
left panel for the legend. In the left column we show the velocity $v(t)$, the
right column depicts the position $x(t)$. The simulations are described in
section \ref{sec:results} below.}
\label{fig:trajectories}
\end{figure*}

After a hard velocity reversal a new waiting time $\tau$ is generated and the
deterministic motion is continued until the next hard velocity reversal. If
during the period $\tau$ a soft velocity reversal occurs at the points of
maximal distance from the origin, the waiting time continues to be counted,
that is, the waiting period in between two hard velocity reversals remains
unaffected by soft reversals. Therefore, the system evolves deterministically
between the random velocity reversals. We expressly note that the velocity
reversals only change the sign of the velocity thus keeping the system
conservative with constant energy (\ref{eq:conservation}). We here assume
each waiting time event leads to a hard velocity reversal with unit
probability.\footnote{We could also assume that velocity reversals occur
with probability $\frac{1}{2}$.} In phase space the particle makes a jump
from the point $(x,v)$ to the point $(x,-v)$ on the orbit determined by
(\ref{eq:conservation}), without change in the direction of motion along
the orbit, see figures \ref{fig:phasespace} and \ref{fig:trajectories}.

\section{Trajectories and mapping onto position and velocity distributions}
\label{trajs}

Figure~\ref{fig:phasespace} schematically depicts the motion of a particle
described by equation (\ref{eq:newton}) along with its velocity reversals in
phase space. Due to energy conservation (clock wise) motion takes place on
the closed $\mathcal{E}=\mathrm{const}$ orbit, which is depicted by the dotted
line. Note that for clarity the solid line representing the sample trajectory is
drawn with a different radius. The points labelled $\mathrm{s}$ and $\mathrm{f}$ indicate the
starting and final positions respectively. Points $1$ and $1'$ as well as
$2$ and $2'$, connected with dashed lines, illustrate the hard velocity reversals.
During these reversals a particle changes its velocity from unprimed to primed
numbers, while the position remains the same. Figure \ref{fig:trajectories}
demonstrates some sample time evolutions of the velocity $v(t)$ (left column)
and position $x(t)$ (right column) of a test particle for various model
parameters and potential types.

\begin{figure}
\centering
\includegraphics[width=10cm]{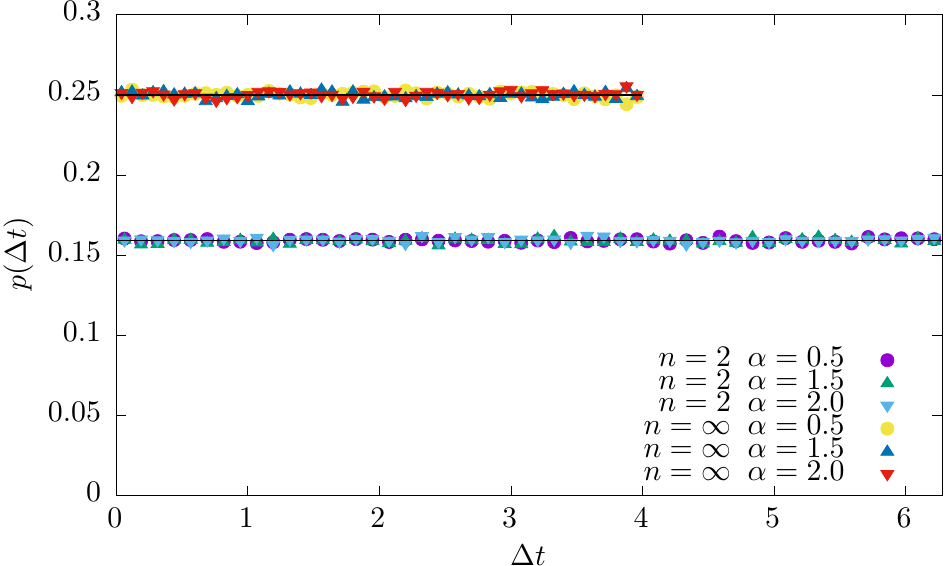}
\caption{Accumulated time shift distribution $p(\Delta t)$, see equation~(\ref{delta_t}), for $n=2$ and $n=\infty$
with various values of $\alpha$. The solid lines represent $1/2\pi$ and $1/4$, which
are uniform densities over $[0,T]$ for $n=2$ ($T=2\pi$) and $n=\infty$ ($T=4$).}
\label{fig:phase}
\end{figure}

Every hard velocity reversal is associated with advancing the solution $x(t)$ of
equation (\ref{eq:newton}) by a shift $\delta$
\begin{equation}
\delta=\left\{
\begin{array}{lcccl}
2\mathcal{T}(|x(t)|,\xm) && \mbox{if} &&
x(t)\times v(t) > 0\\ T-2\mathcal{T}(|x(t)|,\xm) && \mbox{if} && x(t)\times
v(t) <0
\end{array}
\right.,
\label{eq:timeshift}
\end{equation}
where $\mathcal{T}(|x(t)|,\xm)$ is the time needed to travel from $|x(t)|$ to the
reversal point $\xm$,
\begin{equation}
\mathcal{T}(|x(t)|,\xm)=\int_{|x(t)|}^{\xm} \frac{dx}{\sqrt{2[\mathcal{E}-V(x)]}}.
\label{eq:time}
\end{equation}
$T$ is the period of the motion given in relation (\ref{eq:period}). The time shift
(\ref{eq:timeshift}) assures that $x(t)=x(t+\delta)$ and $v(t)=-v(t+\delta)$, which
in turn guarantee energy conservation. Numerical simulations of the model confirm
that at sufficiently long time $t$ the accumulated time shift
\begin{equation}
\Delta t=\sum_{i=0}^N \delta_i \;\mod\; T
\label{delta_t}
\end{equation}
is uniformly distributed on the interval $[0,T)$. The upper summation bound $N$ is
defined by the condition
\begin{equation}
\sum_{i=0}^{N-1}\tau_i\leqslant t\leqslant\sum_{i=0}^{N}\tau_i,
\end{equation}
where the $\tau_i$ are the independent identically distributed random waiting
times defined by the distribution (\ref{eq:tauasymptotics}). The observation that
$\Delta t$ in expression (\ref{delta_t}) is uniform, $U([0,T))$, is shown in
figure \ref{fig:phase} for $n=2$ and $n=\infty$.

Due to the velocity reversals the phase space co-ordinates $x(t)$ and $v(t)$ are no
longer deterministic but they become random variables. Moreover, position $x$ and
velocity $v$ are not independent but are linked by the energy conservation
constraint (\ref{eq:conservation}). Therefore, the velocity density $p(v)$ can be
used to calculate the position distribution $p(x)$ through
\begin{equation}
p(x)=\frac{|x|^{n-1}}{\sqrt{2\left[\mathcal{E}-V(x)\right]}}\times p\left(v=\pm
\sqrt{2\left[\mathcal{E}-V(x)\right]}\right).
\label{eq:transformation}
\end{equation}
Analogously, $p(v)$ can be obtained from $p(x)$ through
\begin{eqnarray}
\nonumber
p(v)&=&\left[n\left(\mathcal{E}-v^2/2\right)\right]^{1/n-1}|v|\\
&&\times p\left(x=\pm\left[n\left(\mathcal{E}-v^2/2\right)\right]^{1/n}\right).
\label{eq:transformation2}
\end{eqnarray}
In figure \ref{fig:randomness} we demonstrate that on top of the soft velocity
reversals at the zero-velocity points, even for \textit{identical initial conditions} hard
reversals still maintain the random nature of the conservative L{\'e}vy walk
motion.
Consequently, after a finite time trajectories corresponding to the \textit{same initial conditions} are becoming disparate, thus randomizing velocity $v(t)$ and position $x(t)$.

The cases $n=1$ and $n=2$ are fully traceable analytically. For $n\geqslant3$ one
has to rely on numerical methods. Consequently, we start our considerations with
the case $n=2$ corresponding to the conservative L{\'e}vy walk harmonic oscillator.
Consecutively, we proceed with the case $n=1$ and, finally, we consider the cases
$n=4$ and $n=\infty$ numerically.

In the specific $n=2$ case the relation (\ref{eq:newton}) describes the harmonic
oscillator, for which
\begin{equation}
x(t)=\xm \sin(t+\varphi),
\label{eq:n2sol}
\end{equation}
where $\varphi$ is the initial phase. Without loss of generality it can be
assumed that $\varphi=0$, i.e., $x(0)=0$ and $v(0)=v_{\mathrm{max}}$. The
first velocity reversal at $t_1$ introduces the extra phase shift $\delta$ to
equation (\ref{eq:n2sol}), which is equal to
\begin{equation}
\delta = \pi-2 t_1.
\end{equation}
The phase shift can be calculated from the condition
\begin{equation}
\sin(t_1)=\sin(t_1+\delta),
\end{equation}
which is fulfilled for $\delta=0$ or $\delta=\pi-2t_1$, as $\sin(u)=\sin(w)$ has
two series of main solutions, $u=w$ and $u=\pi-w$. Only the latter solution
assures that $v(t_1)=-v(t_1)$. Finally, for $n=2$ the phase shift fulfils the
recursion
\begin{equation}
\left\{\begin{array}{ccl}\delta_0&=&0\\\delta_n&=&\pi-2t_n-\delta_{n-1}\end{array}
\right.,
\label{eq:n2recursion}
\end{equation}
where $t_n$ are time instants of velocity reversal. These time instants are given
by
\begin{equation}
t_n=t_{n-1}+\tau
\end{equation}
with $t_0=0$.

For $\alpha>1$ (or for the exponential density) when the mean time $\tau$ between
reversals is finite, from numerical simulations we conclude that at sufficiently
long times $\Delta t$, see equation~(\ref{delta_t}), is uniformly distributed on $[0,2\pi)$, see
figure~\ref{fig:phase}. 
\rev{Analogously for $\alpha<1$ (see below) the distribution $p(\Delta t)$ also
becomes uniform, however, due to the long memory of the initial conditions a
significantly longer time is required as compared to the case $\alpha>1$.}
Therefore, both velocity and position are distributed
according to the arcsine laws
\begin{equation}
p(v)=\frac{1}{\pi\sqrt{1-v^2}}
\label{eq:n2velocity}
\end{equation}
and
\begin{equation}
p(x)=\frac{1}{\pi\sqrt{1-x^2}},
\label{eq:n2position}
\end{equation}
where we interpret velocity and position as $v(t)/\vm$ and $x(t)/\xm$. Above
results follow from the transformation of variables
\begin{equation}
v(t)/\vm=\cos(t+\delta \;\mod\; 2\pi),
\end{equation}
and
\begin{equation}
x(t)/\xm=\sin(t+\delta \;\mod\; 2\pi),
\end{equation}
where $(t+\delta \;\mod\; 2\pi)$ is uniformly distributed on $[0,2\pi)$, that is,
$U([0,2\pi))$. This behaviour is indeed corroborated by our numerical analysis,
see below.

A different situation occurs for $\alpha<1$, when the mean waiting time between
hard velocity reversals diverges. In this situation, even at appreciably long
times discrete \rev{slowly decaying} peaks appear in the distributions $p(x)$ and $p(v)$, see \cite{dybiec2017levy}. These
correspond to deterministic motion events without hard velocity reversal, that is,
to the solution of equation (\ref{eq:newton}) with below initial conditions (\ref{eq:icx})
and (\ref{eq:icv}). At the time instant when $p(x)$ and $p(v)$ are evaluated, the
trajectories corresponding to the peaks will be at exactly the same position with
exactly the same velocity. With increasing time the height of these peaks decreases
as the likelihood for hard velocity reversals increases. 
\rev{Finally, in the limit $t\to\infty$, in analogy to the case $\alpha>1$, the
peaks also disappear.}
Of course, these peaks can
be completely eliminated by taking a random initial condition on the constant energy
orbit.

A similar effect is observed for $V(x)=|x|$, in this case $v(t)$ is piecewise
linear and $x(t)$ is piecewise parabolic. Assuming, analogously to the harmonic
oscillator, that the velocity reversal introduces an extra phase shift to $x(t)$
and $v(t)$ leading to uniform distribution of $(t+\delta\;\mod\; T)$ on $[0,T)$,
transformation of variables shows that the velocity is uniformly distributed
over $[-\vm,\vm]$,
\begin{equation}
p(v)=\frac{1}{2\vm}.
\label{eq:n1velocity}
\end{equation}
Furthermore, using relations (\ref{eq:transformation}) and (\ref{eq:n1velocity})
the density $p(x)$ can be calculated as
\begin{equation}
p(x)=\frac{1}{4\sqrt{\mathcal{E}(\mathcal{E}-|x|)}},
\label{eq:n1position}
\end{equation}
where $x\in[-\xm,\xm]=[-\mathcal{E},\mathcal{E}]$. Analogously, expression
(\ref{eq:n1position}) can be calculated using the fact that $x(t)$ is piecewise
parabolic. Our numerical results thus confirm the above assumption of the
uniform distribution of $(t+\delta\;\mod\; T)$.

\section{Numerical analysis}
\label{sec:results}

\begin{figure}
\centering
\includegraphics[width=8cm]{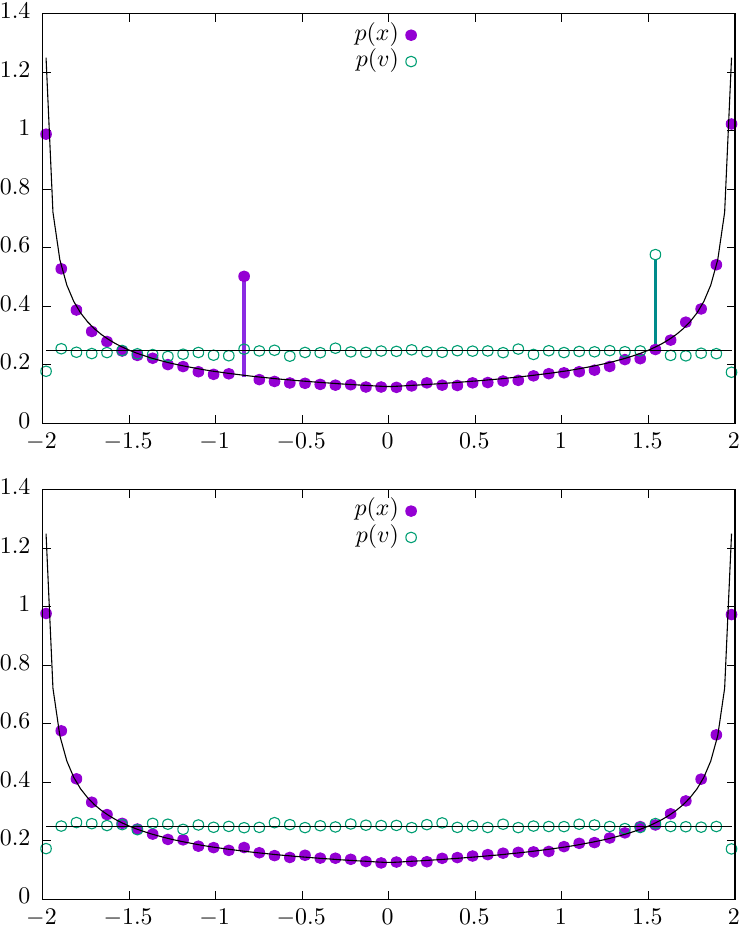}\\
\caption{Probability densities $p(x,t=10^3)$ and $p(v,t=10^3)$ for $V(x)=|x|$ with $\alpha=0.5$
(top) and $\alpha=1.5$ (bottom). Solid lines are given by equations
(\ref{eq:n1velocity}) and (\ref{eq:n1position}). The initial conditions are
adjusted in such a way that $\vm=\xm$, see relations (\ref{eq:icx}) and
(\ref{eq:icv}).}
\label{fig:n1}
\end{figure}

\begin{figure}
\centering
\includegraphics[width=8cm]{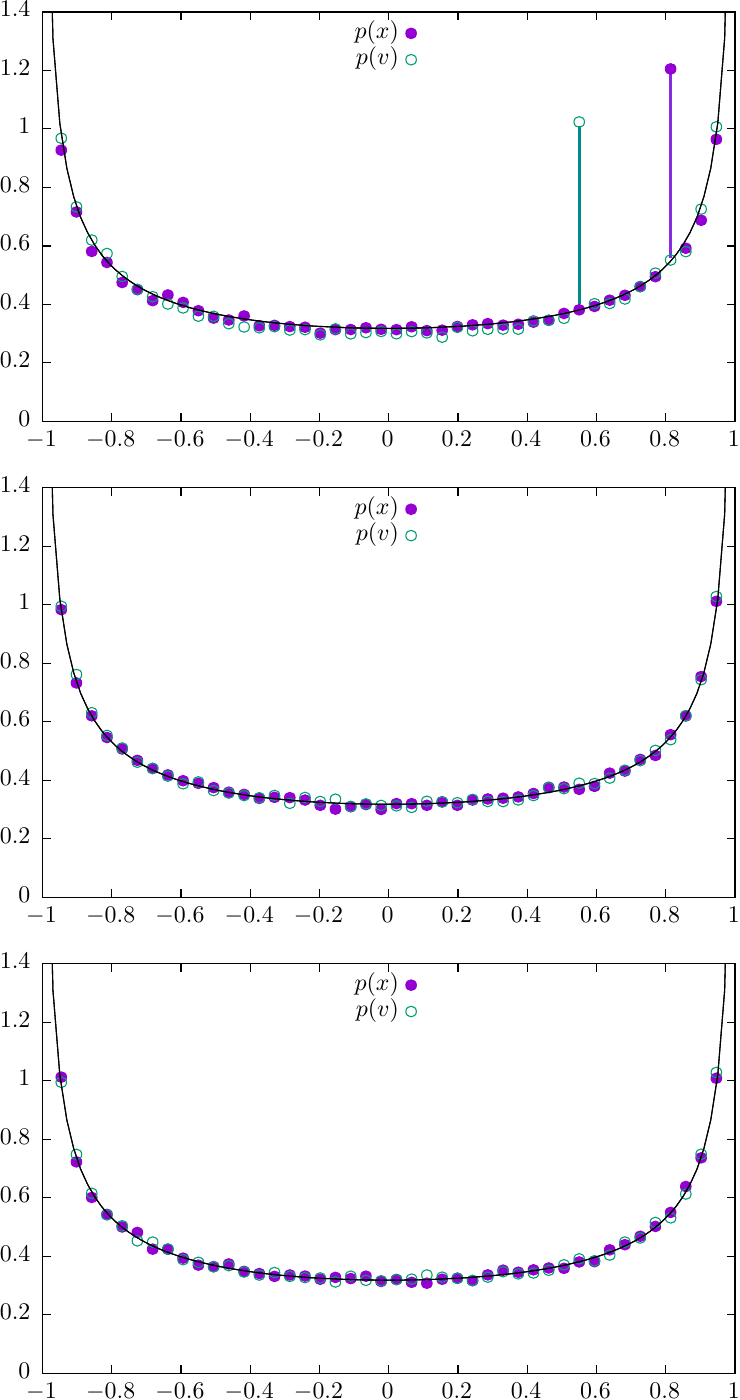}\\
\caption{Probability densities $p(x,t=10^3)$ and $p(v,t=10^3)$ for $V(x)=x^2/2$ with $\alpha=0.5$
(top) and $\alpha=1.5$ (middle). We also show the case of an exponential waiting
time density with the scale factor $\lambda=1$ (bottom). Solid lines correspond to
the values in equations (\ref{eq:n2velocity}) and (\ref{eq:n2position}).}
\label{fig:n2}
\end{figure}

\begin{figure}
\centering
\includegraphics[width=8cm]{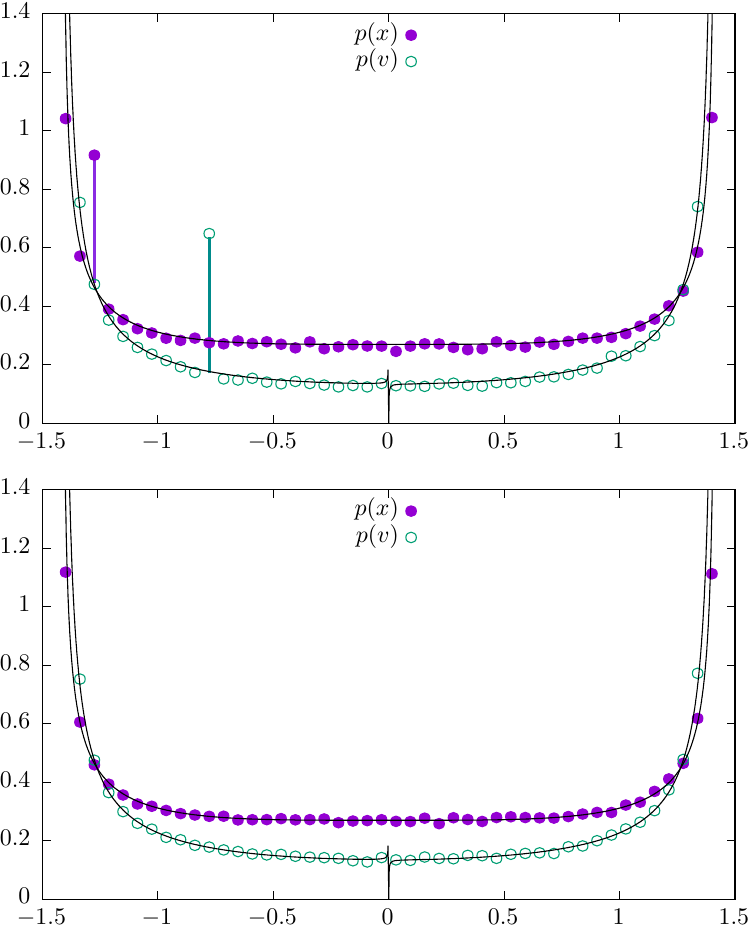}\\
\caption{Probability densities $p(x,t=10^3)$ and $p(v,t=10^3)$ for $V(x)=x^4/4$ with $\alpha=0.5$
(top) and $\alpha=1.5$ (bottom). The solid lines are calculated using Wolfram
Mathematica.}
\label{fig:n4}
\end{figure}

\begin{figure}
\centering
\includegraphics[width=8cm]{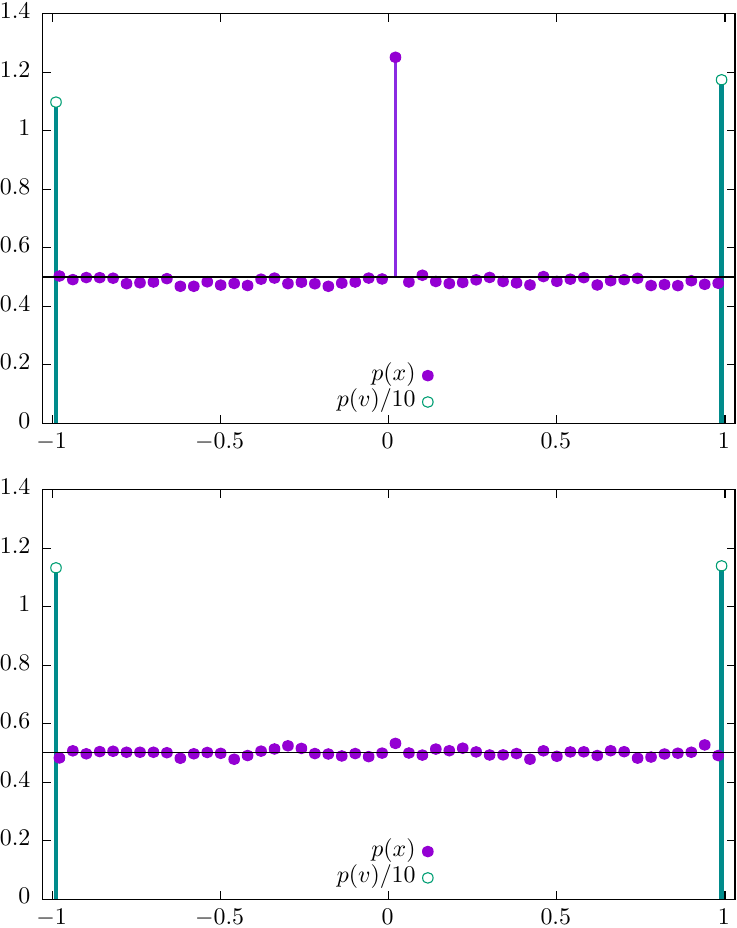}\\
\caption{Probability density $p(x,t=10^3)$ and $p(v,t=10^3)$ for infinite rectangular potential well with $\alpha=0.5$ (top panel) and $\alpha=1.5$ (bottom panel).}
\label{fig:infty}
\end{figure}

As the motion in our conservative phase space is restricted to the total energy
surface $\mathcal{E}=\frac{1}{2}v^2+V(x)=\mathrm{const}$ the velocity $v(t)$
changes between the maximal values $-\vm$ and $\vm$. Similarly, the position
$x(t)$ is restricted to $[-\xm,\xm]$, in contrast to thermally activated
motion which can reach arbitrary values of the phase space co-ordinates, with
the respective Boltzmann weights. For clarity of presentation the initial
conditions are chosen such that the numerical values of $\xm$ and $\vm$ are
identical and the resulting $p(x)$ and $p(v)$, for fixed $n$, densities have the same support.
For our case of unit mass and $n\neq 2$, the condition $\xm=\vm$ uniquely
determines the total energy $\mathcal{E}$, which in turn can be used to
calculate $\xm$ and $\vm$. For $n=2$ the closed orbit is a circle, and thus
the relation $\xm=\vm$ always holds. As the initial condition any point on the
$\mathcal{E}=\mathrm{const}$ orbit can be chosen---we note that some of the
observed results are sensitive to the initial conditions, as discussed below.
We made the explicit choice
\begin{equation}
x(0)=0
\label{eq:icx}
\end{equation}
and
\begin{equation}
v(0)=
\left\{
\begin{array}{lll}
(2/n)^{1/(2-n)}&\mbox{for}&n\neq2\\
1&\mbox{for}&n=2\\
\end{array}
\right..
\label{eq:icv}
\end{equation}
Different initial conditions will introduce an initial phase shift to the
solutions $x(t)$ and $v(t)$. \rev{In principle, we would expect that for
the case $\alpha>1$ fast mixing of the system occurs such that the long time
properties are independent of the initial conditions. For $0<\alpha<1$,
however, when the mean waiting time $\langle\tau\rangle$ diverges, the initial
conditions should be visibly more persistent. In our finite-time simulations,
we indeed observe a fast decay of the initial peaks in the position and
velocity distributions for $\alpha>1$, while for $0<\alpha<1$ they are slower.
Yet we do not see a clear change of the generic behaviour when $\alpha$ crosses
the value of unity, see figures \ref{fig:n1}, \ref{fig:n2}, \ref{fig:n4}, and
\ref{fig:infty}.}

As evidenced in figure~\ref{fig:slope} this decay is actually of power-law form
whose scaling exponent depends on the exponent of the waiting time density. It
is tempting to assume that the decay of the peak height is directly linked to the
scaling exponent $\alpha$ of the waiting time density, in the spirit of ``The
single big jump principle'' \cite{vezzani2018single}. However,
the results listed in table~\ref{tab:exponents} appear not fully conclusive.
For random
initial conditions on the constant energy orbit, in contrast to fixed initial conditions, even for $\alpha<1$
there are no additional peaks in the densities $p(x)$ and $p(v)$ as the peaks
are washed out. 

For the general potential (\ref{eq:potential}) with exponent $n$ the period of the
deterministic motion is given by expression (\ref{eq:period}). The values of the
periods $T$ corresponding to different $n$ with initial conditions (\ref{eq:icx}) and
(\ref{eq:icv}) (or, more precisely, with the total energy $\mathcal{E}$ determined
by the initial conditions) are listed in table \ref{tab:periods}. For $n=4$,
these values were used to obtain $p(x)$.

\begin{figure}
\centering
\includegraphics[width=8cm]{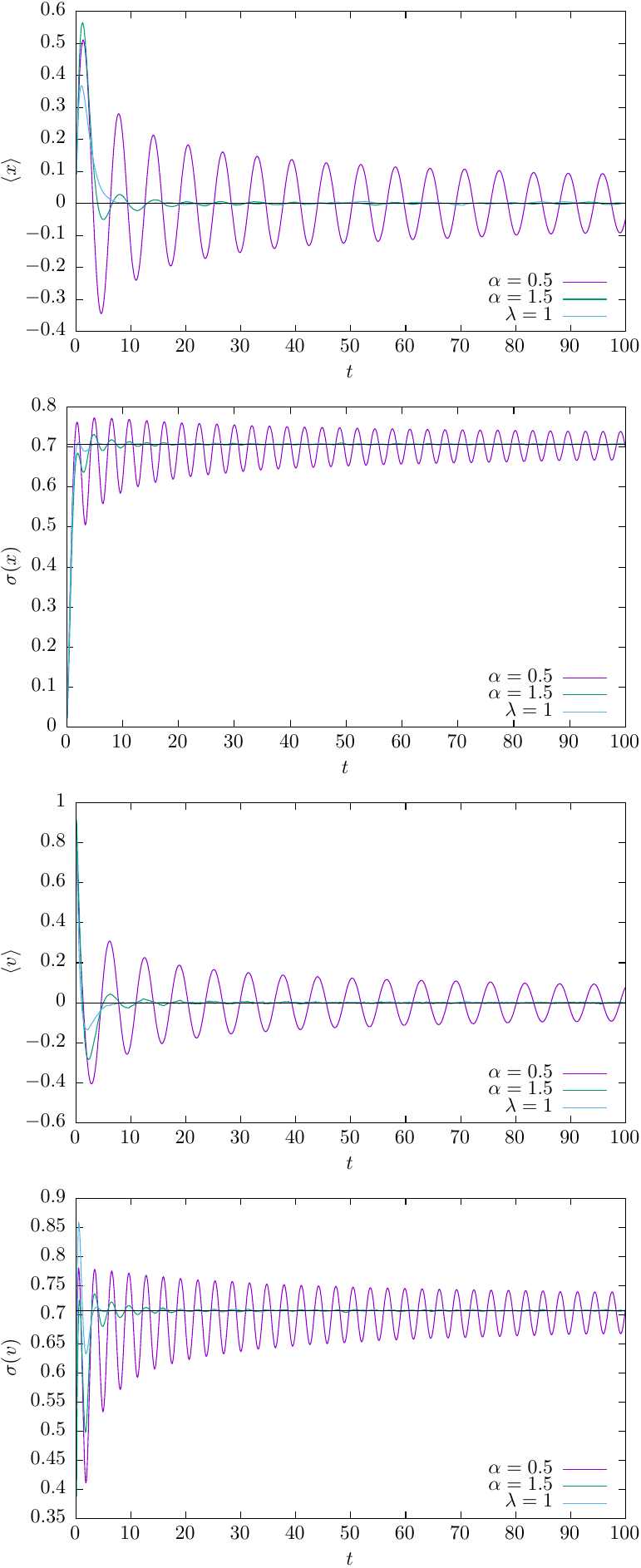}\\
\caption{Case $n=2$, from top to bottom: mean position $\langle x(t)\rangle$,
standard deviation $\sigma(x(t))$ of the position co-ordinate, mean velocity
$\langle v(t)\rangle$, and standard deviation $\sigma(v(t))$ of the velocity.
Solid lines represent the theoretical values of the averages and standard
deviations given in table \ref{tab:stdev}. The initial conditions (\ref{eq:icx})
and (\ref{eq:icv}) assure that $\vm=\xm$.}
\label{fig:n2-time}
\end{figure}

\begin{table}
\vspace*{0.4cm}
\centering
\begin{tabular}{c|c}
\hline
$n$ & $T$\\[2pt]\hline\hline
1&8\\
2&$2\pi\approx6.283$\\
4 & $\sqrt{\pi}\Gamma\left[\frac{5}{4}\right]/\Gamma\left[\frac{3}{4}\right]
\approx5.244$\\
$\infty$ & $4$\\\hline
\end{tabular}
\vspace*{0.4cm}
\caption{Periods $T$ of the periodic motion in potentials $V(x)=|x|^n/n$, see
relation (\ref{eq:period}), for various values of the power $n$, with initial
conditions (\ref{eq:icx}) and (\ref{eq:icv}).}
\label{tab:periods}
\end{table}

\begin{table}
\vspace*{0.4cm}
\centering
\begin{tabular}{c|c|c}
\hline
$n$ & $\sigma(x)$ & $\sigma(v)$\\\hline \hline
1 & $\sqrt{32/15}\approx1.461$ & $\sqrt{4/3}\approx1.155$\\
2 & $\sqrt{1/2}\approx0.707$ & $\sqrt{1/2}\approx0.707$\\
4 & 0.956 & 1.155\\
$\infty$ &  $\sqrt{1/3}\approx0.577$ & 1 \\ \hline
\end{tabular}
\vspace*{0.4cm}
\caption{Asymptotic values of the standard deviations $\sigma(x)$ and $\sigma(v)$
of the position and velocity co-ordinates for periodic motion in the potentials
(\ref{eq:potential}), for various values of $n$ and initial conditions
(\ref{eq:icx}) and (\ref{eq:icv}).}
\label{tab:stdev}
\end{table}

In the numerical evaluation for the case $n=4$ with Wolfram Mathematica equation
(\ref{eq:newton}) was used to construct the mapping $[0,T)\ni t\mapsto
x(t)$. Then, using the constructed map and the assumption that $(t+\delta\;\mod\; T)$
is uniform on $[0,T)$ the probability density $p(x)$ was obtained numerically by
transformation of variables. In the final step, $p(x)$ was transformed into $p(v)$
using relation~(\ref{eq:transformation2}).
With increasing steepness $n$ the potential wells become almost flat in the
vicinity of the origin. This implies that in that region there is practically
no external force acting on the test particle. Consequently, close to the origin
for large $n$ the velocity is practically constant, see the third panel from the
top in figure \ref{fig:trajectories}. Therefore, the procedure based on the $[0,T)
\ni t\mapsto v(t)$ mapping cannot be successfully applied to determine $p(v)$ when
$x(t)\approx0$. To calculate the density $p(v)$ a transformation of variables $x\to
v$ has to be used, see equation (\ref{eq:transformation2}).
Nevertheless, despite being more robust than $[0,T)\ni t\mapsto v(t)$ mapping, even
the numerical transformation of variables results in some numerical instabilities,
see figure~\ref{fig:n4}.

\subsection*{The densities $p(x)$ and $p(v)$}

The conservative random walk model based on equations (\ref{eq:newton}) and
(\ref{eq:potential}) as well as random, hard velocity reversals was studied
by Monte Carlo simulations. Multiple realisations were simulated with the
velocity Verlet algorithm \cite{press1992}. Individual sample trajectories of finite length
are displayed in figure \ref{fig:trajectories} for various systems parameters.
In particular, for finite time $t$, we show the difference between power-law waiting times with
finite and infinite mean waiting time, with exponential waiting times.

From the ensemble of realisations the densities $p(x)$ and $p(v)$ were constructed
and compared with the theoretical predictions. Figure \ref{fig:n2-time} shows the
oscillatory dynamics with a converging envelope of the
mean values and standard deviation of the position $x(t)$ and the velocity $v(t)$
for the harmonic potential, $n=2$. Due to the symmetry of the setup, even for an
asymmetric initial condition the densities for position and velocity converge to
symmetric forms. Therefore the average values asymptotically tend to zero. As
shown in figure \ref{fig:slope} the decay of the envelope of this dynamics is of
power-law form.
The standard deviations converge to their asymptotic values given in table
\ref{tab:stdev}. 
For diverging mean waiting time for hard velocity reversals
corresponding to $0<\alpha<1$, clear periodicities in the mean and standard
deviations of position and velocity are visible. 
\rev{Nevertheless, as discussed in \ref{app1}, the amplitude of the periodicity
decays in time with a characteristic time scale depending on the value of the
exponent $\alpha$, see figure~\ref{fig:slope} and table~\ref{tab:exponents}.}
The observed periodicity agrees
with the period of the motion for $\langle x(t)\rangle$ and $\langle v(t)\rangle$,
as provided in table \ref{tab:periods}. The time dependencies of $\sigma(x(t))$
and $\sigma(v(t))$ are characterised by the half periods as given in table
\ref{tab:periods}: $\langle x^2(t)\rangle$ and $\langle v^2(t)\rangle$ indeed
feature half the periodicity of $\langle x(t)\rangle$ and $\langle v(t)\rangle$.
Table \ref{tab:stdev} provides the asymptotic values of the standard deviations.
For $n=1$, $n=2$ and $n=\infty$ they are calculated exactly from the theoretical
densities $p(x)$ and $p(v)$, while for $n=4$ they were obtained numerically from
$p(x)$ and $p(v)$.

Figures \ref{fig:n1}, \ref{fig:n2}, \ref{fig:n4}, and \ref{fig:infty} display 
the finite time distributions $p(x,t)$ and $p(v,t)$ with $\alpha=0.5$, $\alpha=1.5$ and
$\kappa=\sqrt{2}$ as well as $\lambda=1$ (exponential waiting times), respectively,
for increasing values of $n$ at $t=10^3$. Localised peaks in $p(x)$ and $p(v)$ for $\alpha=0.5$
are located at values of $x(t)$ and $v(t)$ corresponding to the deterministic
motion, see equation~(\ref{eq:newton}).
The results for the exponential waiting time distribution ($\lambda=1$) are
analogous to the case with finite mean waiting time corresponding to $1<\alpha \leqslant 2$.
Therefore, we conclude that the model properties are not very sensitive to the
exact shape of the waiting time distribution. \rev{The finiteness or infinity of
the mean waiting time for $1<\alpha<2$ and $0<\alpha<1$, respectively, appears to
only set the rate of convergence to the stationary state.}

The solid lines in figures \ref{fig:n1}, \ref{fig:n2}, \ref{fig:n4}, and
\ref{fig:infty} represent the theoretical curves given by relations
(\ref{eq:n2velocity}) and (\ref{eq:n2position}) as well as (\ref{eq:n1velocity})
and (\ref{eq:n1position}). For the case $n=4$ they are calculated with Wolfram
Mathematica. As mentioned above, the initial conditions are adjusted such that
$\vm=\xm$, as given by equations (\ref{eq:icx}) and (\ref{eq:icv}). Some numerical
instabilities are visible in figures \ref{fig:n4} where the distribution $p(v)$,
calculated from $p(x)$, fluctuates around the zero line. Small velocities are
recorded when $x\approx\pm\xm$. In this region the numerical inversion of $x(t)$
introduces some error, which is responsible for the $p(v)$ fluctuations.

In the limit of $n\to\infty$ the potential $V(x)=|x|^n/n$ becomes comparable to
the infinite deep rectangular potential well, see figure \ref{fig:potential}. In
this case $p(x)$ is uniform and the formula for $p(v)$ reads
\begin{equation}
p(v)=\frac{1}{2}\left[\delta(v-v_0)+\delta(v+v_0)\right],
\end{equation}
which is nicely corroborated in figure \ref{fig:infty}. Note that $v_0$ is set
to $v_0=1$ and $\xm$ is set to $\xm=1$.
For $0<\alpha<1$ isolated \rev{persistent} peaks in the densities $p(x)$
and $p(v)$ are visible, see figure \ref{fig:infty}.

\subsection*{Tails versus central parts of waiting time distributions}

\begin{figure}
\centering
\includegraphics[width=10cm]{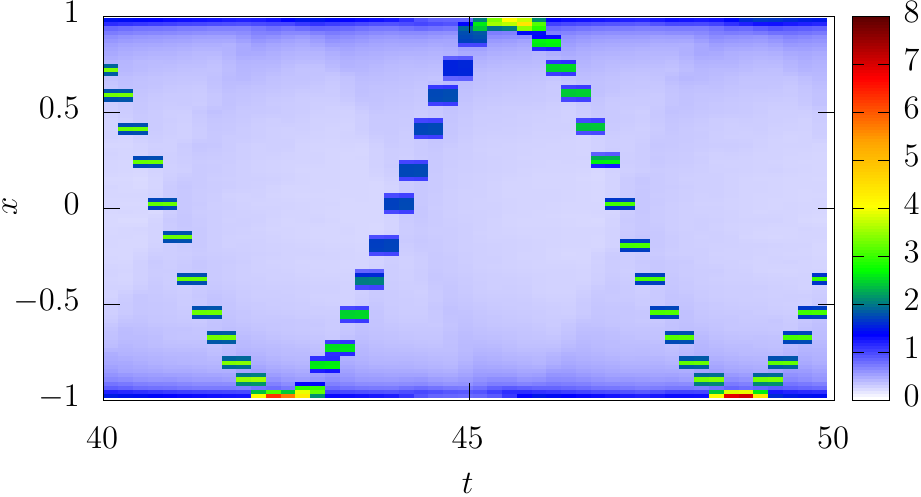}
\caption{Sample time dependent probability density $p(x,t)$ for $\alpha=0.5$,
$n=2$, and initial conditions (\ref{eq:icx}) and (\ref{eq:icv}) The motion
appears almost fully deterministic due to lack of hard velocity reversals.
See also the discussion in section \ref{sec:summary}.}
\label{fig:n2-a05-map}
\end{figure}

\begin{figure}
\centering
\includegraphics[width=8cm]{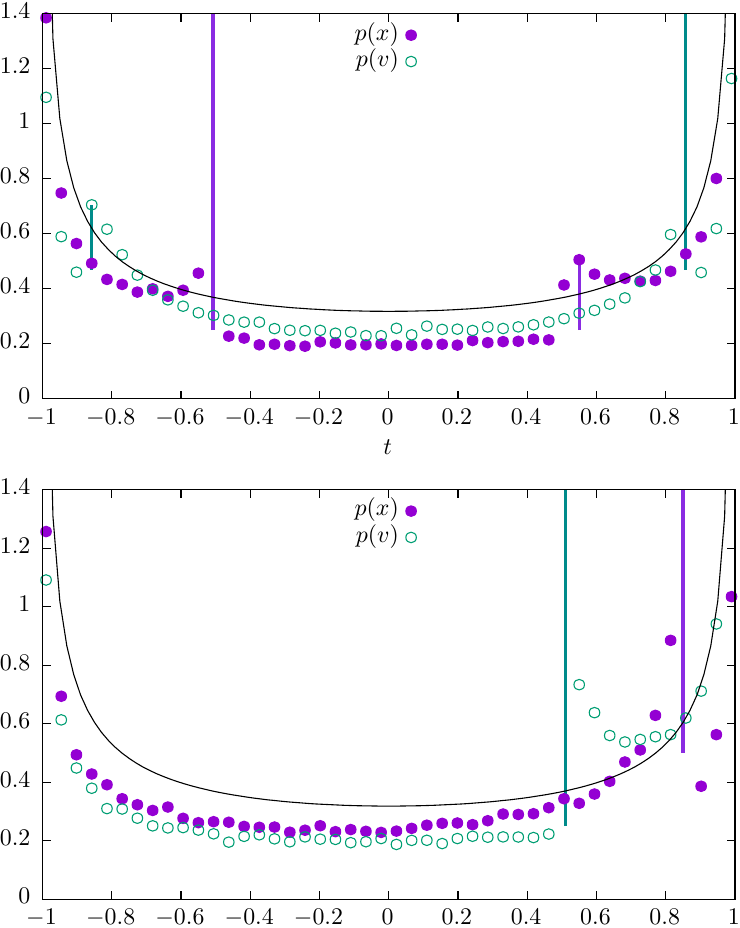}\\
\caption{Distributions of $p(x,t=100)$ and $p(v,t=100)$ for $\alpha=0.3$ with $n=2$ for the initial condition given by Eqs.~(\ref{eq:icx}) and~(\ref{eq:icv}) (top panel) and $v(0)=0$ with $x(0)=\xm$ (bottom panel).}
\label{fig:a03}
\end{figure}

Figure~\ref{fig:n2-a05-map} presents the sample time dependent probability
density $p(x,t)$ as a heat-map. The presented results correspond to $\alpha=0.5$,
$n=2$ for initial conditions (\ref{eq:icx}) and (\ref{eq:icv}). As this case is
part of our discussion for diverging mean waiting time, the dominating \rev{slowly disappearing} maxima of
$p(x,t)$ correspond to the deterministic motion $x(t)$ without hard velocity
reversals. The projection of the maxima is thus given by $\sin(t)$, the solution
of equation (\ref{eq:newton}) for the selected setup. The maximum value of $p(x,t)$
decays with time due to hard velocity reversals, which eventually will occur. The
decay rate is determined by $\alpha$ in the sense that larger values of $\alpha$
lead to more frequent randomising hard velocity reversals. Note the faint shadow
line of the sine function shifted by half a period, $\delta=\pi$. Its origin is
discussed in relation to figure \ref{fig:a03}.
\rev{The case of $\alpha<1$ when the decay of the peaks is slow, should be
contrasted with the case $\alpha>1$: namely, when the mean time between velocity
reversals is finite, the disappearance of the maxima of $p(x,t)$ is fast.}

\begin{figure}
\centering
\includegraphics[width=8cm]{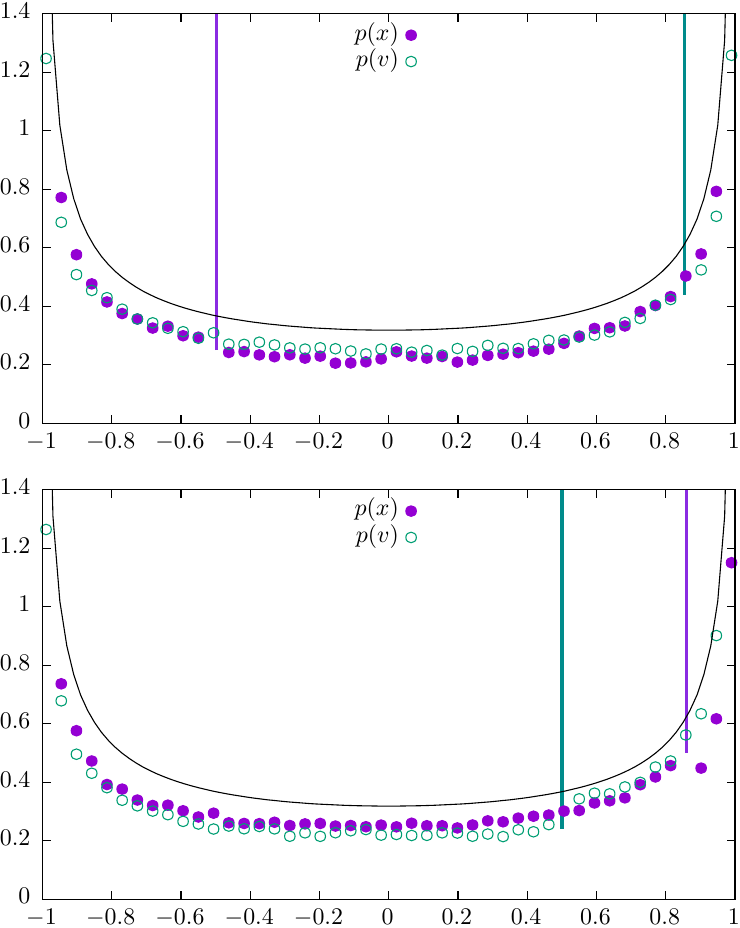}\\
\caption{Distributions of $p(x,t=100)$ and $p(v,t=100)$ for Pareto distributions with $\alpha=0.3$ and $n=2$ for the initial condition given by Eqs.~(\ref{eq:icx}) and~(\ref{eq:icv}) (top panel) and $v(0)=0$ with $x(0)=\xm$ (bottom panel).}
\label{fig:a03-pareto}
\end{figure}

\begin{figure}
\centering
\includegraphics[width=8cm]{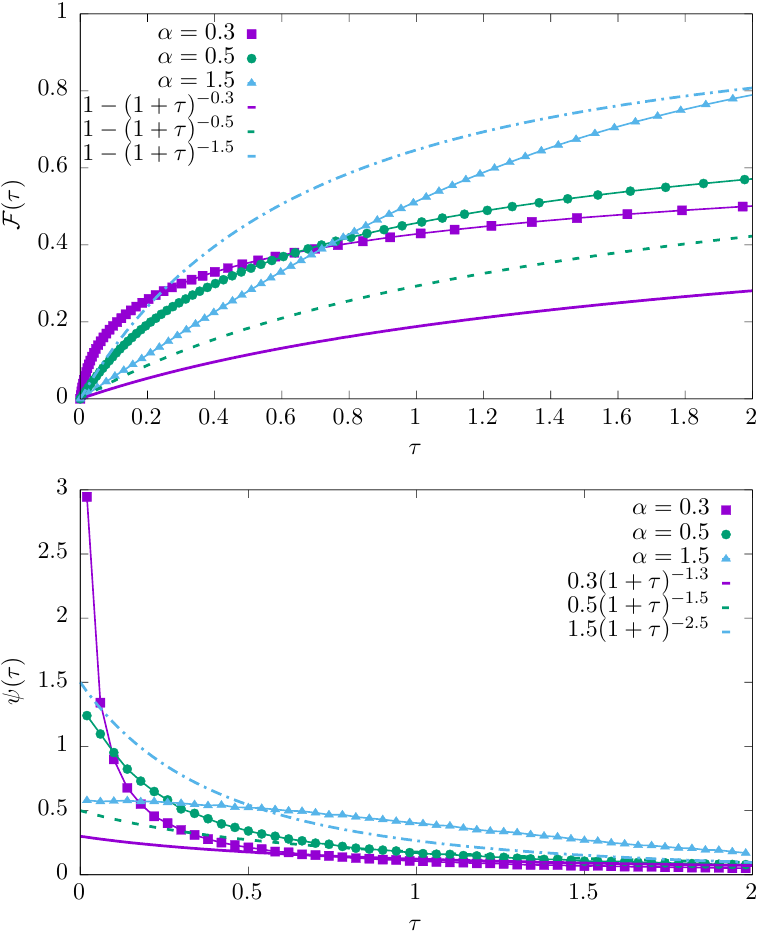}\\
\caption{The associated cumulative distributions $\mathcal{F}(\tau)$ (top panel) and probability densities $\psi(\tau)$  (bottom panel) for $\alpha$-stable and Pareto densities with $\alpha=0.3,0.5,1.5$.}
\label{fig:distributions}
\end{figure}

An interesting dependence is observed for even smaller values of the stability
index $\alpha$, see top panel of figure \ref{fig:a03}.
For $\alpha=0.3$ and up to intermediate
times the distributions $p(x)$ and $p(v)$ exhibit two peaks. The higher dominating
peak corresponds to deterministic motion without hard velocity reversals, which
occur with appreciable probability due to the scale-free nature of the waiting
time distribution. The lower, symmetrically localised peak corresponds to those
trajectories which were immediately reversed. A significant fraction of those
reversed trajectories continues deterministically, thus giving rise to the
secondary \rev{transient} peak in position and velocity distributions. With increasing time the
height of the peaks is decreasing and finally the system reaches its stationary
density. The height of the secondary peak is decaying relatively fast with the
stability index $\alpha$. A relatively faint reminiscence of the secondary peak
can also be observed for stability index $\alpha=0.5$ in figure \ref{fig:n2-a05-map}.
These secondary peaks arise due to immediately reversed trajectories are confirmed
by the cumulative density of waiting times, see figure \ref{fig:distributions},
which shows that for small $\alpha$ there is a significant probability of
(immediate) reversals.

Initial conditions~(\ref{eq:icx}) and (\ref{eq:icv}) represent the situation
when the particle motion is started in the minimum of the potential with the
non-zero initial velocity. In the course of time velocity decreases as the
potential energy grows. Finally, when $x=\xm$ the velocity becomes equal
to 0 and a soft reversal takes place. Such an initial condition results
in the maximal initial velocity.  Consequently, an immediate hard velocity
reversal is well visible and it results in the appearance of secondary peaks in
position and velocity distributions, see top panel of figure~\ref{fig:a03}. In
contrast to (\ref{eq:icx}) and (\ref{eq:icv}) one can assume that a motion
is started at $x=\xm$ with zero initial velocity. On the one hand, both
initial conditions result in the motion along the same orbit. On the other
hand, zero initial velocity prevents immediate velocity reversal, see bottom
panel of figure~\ref{fig:a03} where the secondary peaks in $p(x)$ and $p(v)$
are absent. Nevertheless, the possibility of reversals is manifested by the
extended widths of the primary (without velocity reversals) peaks.

Finally, in order to verify the hypothesis of the immediate reversal we used a
different waiting time distribution, namely, the Pareto density
\begin{equation}
p(\tau)=\alpha(1+\tau)^{-(\alpha+1)},
\label{eq:pareto}
\end{equation}
with $\alpha>0$. The density~(\ref{eq:pareto}) has the same power-law
asymptotics as the applied $\alpha$-stable density, see Eq.~(\ref{eq:charact}).
Figure~\ref{fig:a03-pareto} presents the same results as figure~\ref{fig:a03}
for the Pareto density~(\ref{eq:pareto}). It is clearly visible that
for the Pareto density of wating times there are no secondary peaks,
corresponding to the immediate velocity reversal, compare top panels
of figures~\ref{fig:a03} and~\ref{fig:a03-pareto}. Additionally,
the Pareto distribution also changes the structure of the peaks in the
absence of an initial velocity, that is for $v(0)=0$, see bottom panels
of figures~\ref{fig:a03} and~\ref{fig:a03-pareto}, when they become narrow
in comparison to the case of an $\alpha$-stable waiting time distribution.
Comparison of figures~\ref{fig:a03} and~\ref{fig:a03-pareto} attributes the
structure of the peaks and the possibility of immediate reversal to the central
($\tau\approx0$) part of the waiting time distributions. This is further
analysed in figure \ref{fig:distributions} in which probability densities $\psi(\tau)$ and cumulative densities
\begin{equation}
\mathcal{F}(\tau)=\mbox{Prob}(\tau'\leqslant\tau)=\int_0^\tau \psi(\tau')d\tau'
\end{equation}
corresponding to $\alpha$-stable and Pareto waiting time distributions are
presented. Figure~\ref{fig:distributions} clearly demonstrates that for
small values of $\alpha$ the central parts ($\tau \approx 0$) of the waiting
time distribution contain more probability mass for an $\alpha$-stable density than for a Pareto
distribution. In turn, this indicates that the central part of waiting time
distribution controls the probability of immediate velocity reversal.

\section{Summary and Conclusions}
\label{sec:summary}

L\'evy walks are continuous time random walks with a spatiotemporal coupling
between jump lengths and waiting times. This effects that long jumps are
penalised by long corresponding waiting times, and thus the emerging mean
squared displacement is always finite, in contrast to uncoupled L{\'e}vy
flights. For the latter the diverging variance of the long-tailed jump length
distribution translates into the divergence of the mean squared displacement.
Most frequently the simplest spatiotemporal coupling using a constant speed
is used. In the velocity model \cite{zumofen1993power} this means that the waiting times
determine the time between velocity changes. In such a scenario no external
forces are considered such that the absolute value of the velocity, the speed
$|v|$, is always constant. 
L\'evy walks are
conservative in the sense that their kinetic energy is constant.

\begin{table}
\vspace*{0.8cm}
\centering
\begin{tabular}{p{0.2\columnwidth}||p{0.2\columnwidth}|p{0.2\columnwidth}|
p{0.2\columnwidth}}
\hline
& $n=1$ & $1<n<\infty$ & $n=\infty$ \\[2pt] \hline\hline
& & & \\
$p(\tau) \propto \tau^{-\alpha-1}$
$$(0<\alpha<1)$$
&
$p(v)$ uniform \linebreak\linebreak\linebreak
$p(x)$ U-shaped
\rev{slowly decaying} peaks in both PDFs
&
$p(v)$ U-shaped \linebreak\linebreak\linebreak
$p(x)$ U-shaped
\rev{slowly decaying} peaks in both PDFs
&
$p(v)=\frac{1}{2}[\delta(v-v_0)+\delta(v+v_0)]$ \linebreak\linebreak
$p(x)$ uniform
\rev{slowly decaying} peaks in $p(x)$ only
\\
\hline
& && \\
$p(\tau) \propto \tau^{-\alpha-1}$
$$(1<\alpha\leqslant 2)$$
&
$p(v)$ uniform \linebreak\linebreak\linebreak
$p(x)$ U-shaped
\rev{fast decaying} peaks in both PDFs
&
$p(v)$ U-shaped \linebreak\linebreak\linebreak
$p(x)$ U-shaped
\rev{fast decaying} peaks in both PDFs
&
$p(v)=\frac{1}{2}[\delta(v-v_0)+\delta(v+v_0)]$ \linebreak\linebreak
$p(x)$ uniform
\rev{fast decaying} peaks in both PDFs
\\\hline
\end{tabular}
\vspace*{0.8cm}
\caption{Probability densities for conservative random walks in single well
potentials for different values of the power exponent $n$ of the external
potential as well as for waiting time densities of power law form with
diverging ($0<\alpha<1$) and finite ($1<\alpha \leqslant 2$) mean waiting time. For
exponential waiting times the same behaviour is found as for the case $1<
\alpha \leqslant 2$.}
\label{tab:densities}
\end{table}

Here, we have studied an extension of L\'evy walk processes to cases in which
external forces influence the motion of the test particle. As in the standard
L\'evy walk scenario we assume that the random walker deterministically continues
its motion for a random time. At the renewal time, a velocity reversal occurs. We
call this a hard velocity reversal, as typically the velocity of the particle
at a random instant of time is finite. In contrast to the standard L\'evy walk
model, however, the speed is no longer constant. In the conservative random
walk model adopted here the total energy of the particle consisting of the potential
energy in the external force field and its kinetic energy, is supposed to be a
constant. This necessarily requires that the velocity changes deterministically
and perpetually, according to Newton's second law of motion. This scenario then
also leads to soft velocity reversals at points of maximal distance from the
centre of the potential, when the potential energy assumes its maximum and the
particle velocity is zero. In this conservative random walk model the sole
source of stochasticity is in the velocity reversal, similar to the standard
L\'evy walk model. Moreover, we demonstrated that the relaxation dynamics of
the system visible in the decay of the envelope of $|\langle x(t)\rangle|$ is
of power-law form.

From analytical calculations and numerical results we calculated the velocity
and position distributions for L{\'e}vy walks in single well potentials
of the form $V(x)=|x|^n/n$ type with $n=1$, $2$, $4$, and $\infty$. The
exact solutions we obtained agree perfectly with the shapes of the densities
estimated from Monte Carlo simulations of the underlying motion. For waiting
times with diverging mean (with power law exponent $0<\alpha<1$) the densities
are decorated with distinct peaks, which \rev{slowly} decay as function of time. 
Regardless of the potential type, the peak height decays as a power-law function
of time, with the decay rate depending on the scaling exponent $\alpha$, see
table~\ref{tab:exponents} and Fig.~\ref{fig:decay}.
For more details see the appendix.
These peaks quickly disappear when the mean waiting time is
finite, corresponding to the cases of power law waiting time distributions
with $1<\alpha \leqslant 2$ or exponential waiting time distributions. When
the motion occurs in symmetric single well potentials with $n=2$ or 4, the
distribution of both velocity and position are U-shaped. For the rectangular
potential well both distributions are flat. In the intermediate case $n=1$
the position distribution is U-shaped while the velocity distribution is
flat. The central properties for the different relevant cases are summarised
in table \ref{tab:densities}.

The studied extension of the L{\'e}vy walk scenario provides a possibility
to verify which details of the dynamics are sensitive to the tails and which
are sensitive to the central parts of the waiting time distributions. Namely,
the tail of the waiting time distribution controls the rate of convergence
to the stationary state. Therefore, for $\alpha<1$, it is responsible for
the appearance of the \rev{slowly vanishing} peaks decorating the stationary states.  The width
of the central part of the waiting time distribution around $\tau\approx0$
is responsible for the appearance of secondary peaks in the position and
velocity distributions which originate in the significant possibility of
immediate velocity reversal. Consequently, there are two control mechanisms
that allow one to eliminate secondary peaks. The first possibility is to use
$v(0)=0$ initial condition. The second option is to use a narrower waiting
time distribution. The former scenario diminishes secondary peaks but still
preserves non-zero width of primary peaks. The latter scenario not only
eliminates secondary peaks but also reduces the width of the primary peaks.
\rev{We note again that regardless of the existence or divergence of the mean
waiting time, in the long time limit the peaks disappear.}

The studied model provides a possible generalisation of the L{\'e}vy
walk model which accounts for external forces.  Within this model, in
analogy to free L{\'e}vy walks the system is conservative.  The system
energy consists of kinetic and potential energy.  The studied model can
be contrasted with other, non-conservative extensions of L{\'e}vy walks
\cite{gradenigo2012einstein,burioni2013rare,dechant2016heavy,dechant2015deviations}.

\ack

This project has been supported in part by the Polish National Science
Center grant (2014/13/B/ST2/02014) and by grant ME 1535/6-1 and ME 1535/7-1
from Deutsche Forschungsgemeinschaft. RM thanks the Foundation for Polish
Science for support within an Alexander von Humboldt Polish Honorary Research
Scholarship. Computer simulations were performed at the Academic Computer
Center Cyfronet, Akademia G\'orniczo-Hutnicza (Krak\'ow, Poland)
under CPU grant DynStoch.

\appendix

\section{Further details on the conservative L{\'e}vy walk dynamics}
\label{app1}

\begin{figure}
\centering
\includegraphics[width=12.8cm]{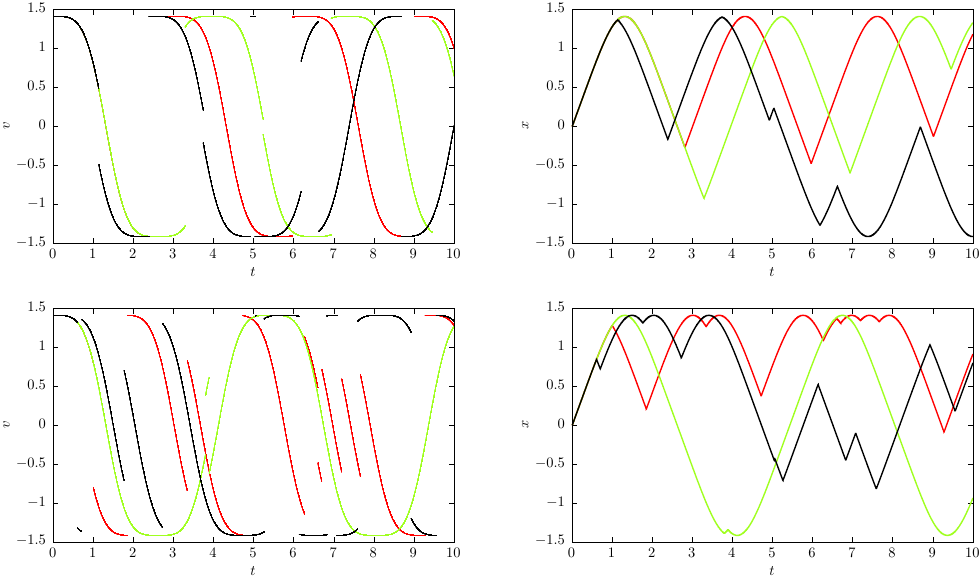}
\caption{Three sample trajectories for $\alpha=0.5$ (top panel) and $\alpha=1.5$
(bottom panel) for $n=4$ with \textit{identical initial conditions}. In the left column we show the velocity $v(t)$,
the right column depicts the position $x(t)$.
The plots evidence the randomisation of sample paths due to hard velocity reversals.
}
\label{fig:randomness}
\end{figure}

Figure \ref{fig:randomness} complements figure \ref{fig:trajectories} and
investigates the role of randomness in the system dynamics. 
In contrast to figure \ref{fig:trajectories},
it presents
three sample trajectories for $\alpha=0.5$ (top panel) and $\alpha=1.5$
(bottom panel) for $n=4$ corresponding to the \textit{same initial conditions}. The
trajectories start with \textit{identical initial conditions} but soon become randomised
due to hard velocity reversals. Figure~\ref{fig:randomness} demonstrates that
with increasing value of $\alpha$ a larger number of hard velocity reversal
is observed.

\begin{figure*}
\centering
\includegraphics[width=12.8cm]{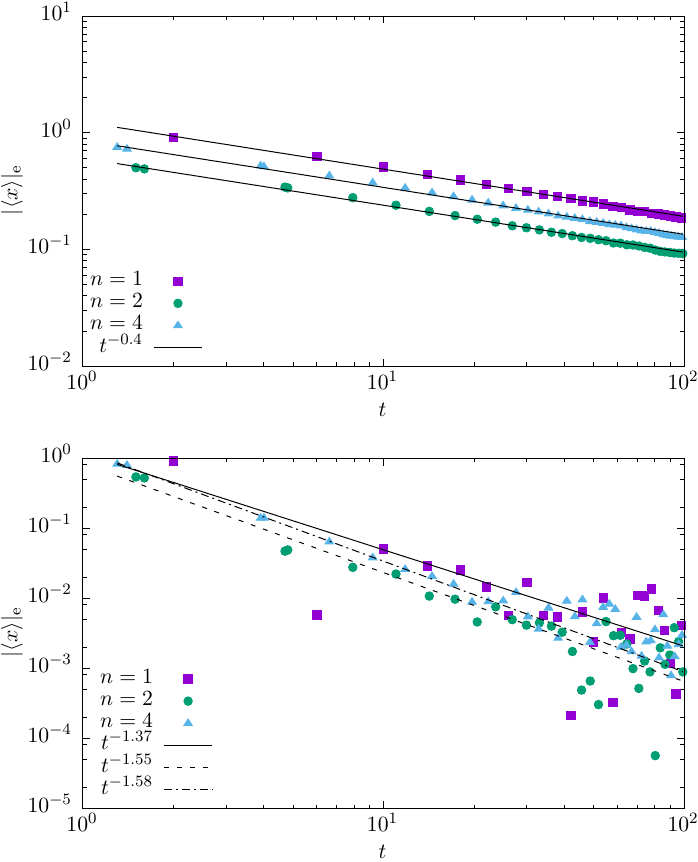}
\caption{Decay of the envelope $|\langle x(t)\rangle|_\mathrm{e}$ representing the
local maxima of $|\langle x(t)\rangle|$, for $\alpha=0.5$ (top panel) and $\alpha=
1.5$ (bottom panel). The scaling exponents for the apparent power-law behaviour are
indicated in the panels.}
\label{fig:slope}
\end{figure*}

In order to quantify how the stationary state is reached we analyse dependence
of $\langle x(t) \rangle$ in more detail. From $\langle x(t)\rangle$ shown in
figure \ref{fig:n2-time} the absolute value $|\langle x(t)\rangle|$ was calculated.
We then determined the envelope $|\langle x(t)\rangle|_\mathrm{e}$ corresponding to
the local maxima of $|\langle x(t)\rangle|$. These envelopes were then fitted by
a power-law. Figure \ref{fig:slope} presents the envelope $|\langle x(t)\rangle|
_\mathrm{e}$ for $\alpha=0.5$ (top panel) and $\alpha=1.5$ (bottom panel) for $n
=1,2,4$. From figure \ref{fig:slope} it is clearly visible that for $\alpha=0.5$
the envelope $|\langle x(t)\rangle|_\mathrm{e}$ decays as a power-law with exponent
$=-0.4$, which is close to the value $0.5$ of $\alpha$, independent of the steepness
of the potential characterised by $n$. For $\alpha=1.5$, in contrast, the power-law
decay is significantly faster with an exponent depending on $n$.

\begin{figure*}
\centering
\includegraphics[width=12.8cm]{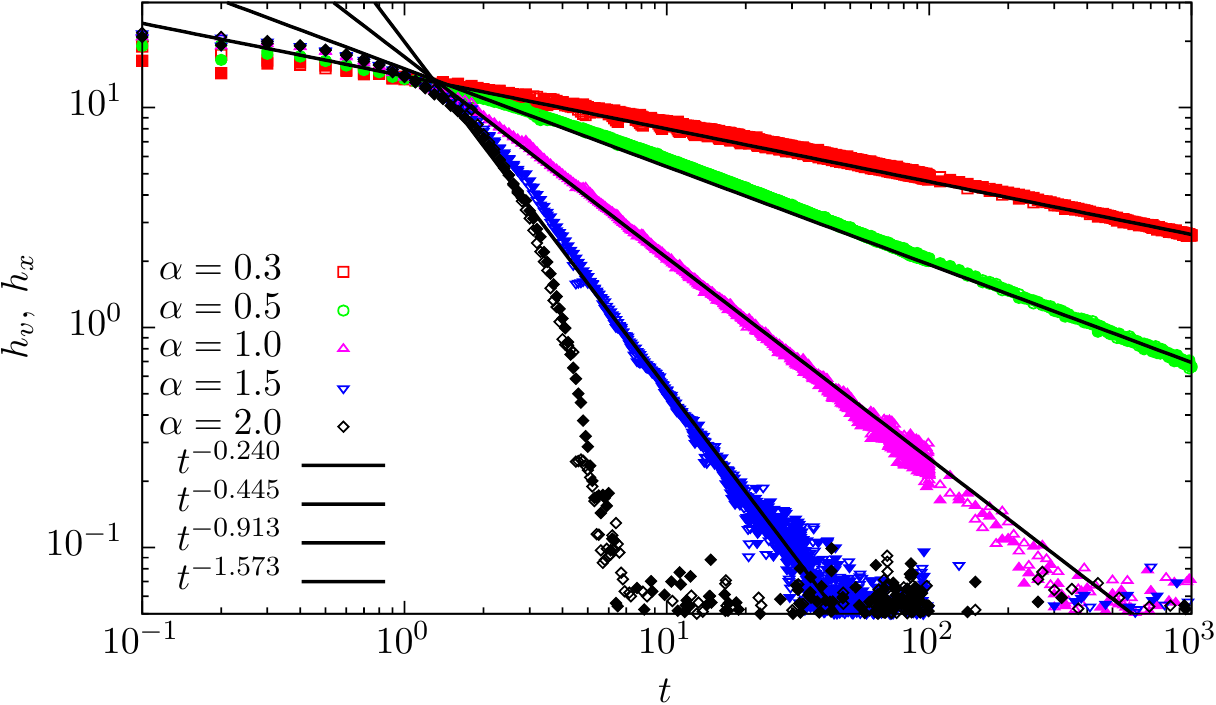}
\caption{Decay of peak height $h_v(t)$ decorating velocity (empty symbols) and position $h_x(t)$ (full symbols) distributions along with fitted slopes, see table~\ref{tab:exponents}.}
\label{fig:decay}
\end{figure*}

Finally, in order to measure the convergence rate to the stationary state we extracted the relative heights $h(t)$ of the deterministic peaks decorating the probability densities. From this data we constructed time series showing peak heights measured from the background given by the density without peaks.
These time series were then used to fit the power-law decay
\begin{equation}
h_v(t) = \beta_v \times  t^{\alpha_v} 
\end{equation}
and
\begin{equation}
h_x(t) = \beta_x \times  t^{\alpha_x} .
\end{equation}
Figure~\ref{fig:decay} presents results for the harmonic potential. Empty symbols show the peak heights in the velocity distribution while full symbols refer to the position distribution.
In both distributions the dependence of the peak height is practically identical.
For $\alpha<2$, the decay rate is of power-law type.
The scaling exponent characterising the decay is very sensitive to the value of the stability index $\alpha$, see table~\ref{tab:exponents}.
For $\alpha<1$ the decay is very slow, consequently peaks decorating the time dependent densities take a very long time to disappear. 
In contrast, for $1<\alpha<2$ the decay still seems to be of power-law type but it is fast enough to diminish the peaks practically quite fast.
\rev{Therefore, on the one hand, regardless of value of $\alpha$ there is the same mechanism responsible for the decay of memory about initial conditions. On the other hand, the observed time scales strongly depend on $\alpha$.}
Interestingly, the exponents describing the decay are weakly  sensitive to the potential type, see table~\ref{tab:exponents}.

\begin{table}
\vspace*{0.4cm}
\centering
\begin{tabular}{c|c||c||c} 
 $n$ & $\alpha $ & $\alpha_v \pm \Delta \alpha_v$  & $\alpha_x \pm \Delta \alpha_x$ \\ \hline \hline 
 1	&	0.3	&	-0.237	$\pm$	0.001 &	-0.238	$\pm$	0.001	\\
1	&	0.5	&	-0.441	$\pm$	0.001 &	-0.441	$\pm$	0.001	\\
1	&	1	&	-0.929	$\pm$	0.003 &	-0.936	$\pm$	0.003	\\
1	&	1.5	&	-1.684	$\pm$	0.009	&	-1.647	$\pm$	0.017	\\
\hline 
2	&	0.3	&	-0.239	$\pm$	0.001 &	-0.240	$\pm$	0.001	\\
2	&	0.5	&	-0.444	$\pm$	0.001 &	-0.445	$\pm$	0.001	\\
2	&	1	&	-0.914	$\pm$	0.003 &	-0.913	$\pm$	0.003	\\
2	&	1.5	&	-1.561	$\pm$	0.013	&	-1.573	$\pm$	0.012	\\
\hline
4	&	0.3	&	-0.238	$\pm$	0.001 &	-0.243	$\pm$	0.001	\\
4	&	0.5	&	-0.439	$\pm$	0.001 &	-0.443	$\pm$	0.001	\\
4	&	1	&	-0.898	$\pm$	0.004 &	-0.901	$\pm$	0.003	\\
4	&	1.5	&	-1.394	$\pm$	0.027	&	-1.621	$\pm$	0.007	\\
\hline
6	&	0.3	&	-0.223	$\pm$	0.002 &	-0.231	$\pm$	0.001	\\
6	&	0.5	&	-0.440	$\pm$	0.002 &	-0.442	$\pm$	0.001	\\
6	&	1	&	-0.905	$\pm$	0.004 &	-0.921	$\pm$	0.003	\\
6	&	1.5	&	-1.298	$\pm$	0.041	&	-1.632	$\pm$	0.009	\\

\end{tabular}
\vspace*{0.4cm}
\caption{Exponents characterising decay of peaks decorating velocity ($\alpha_v$) and positions ($\alpha_x$) distributions as a function of the stability index $\alpha$ and the potential steepness $n$.}
\label{tab:exponents}
\end{table}

\end{document}